\documentstyle[version2,preprint,aps]{revtex}
\begin{document}
\tightenlines
\begin{title}
Order $\alpha^2$ corrections to the decay rate of orthopositronium
\end{title}
\author{G.S. Adkins}
\begin{instit}
Franklin and Marshall College, Lancaster, PA 17604
\end{instit}
\moreauthors{R.N. Fell}
\begin{instit}
Brandeis University, Waltham, MA 01742
\end{instit}
\moreauthors{J. Sapirstein}
\begin{instit}
Department of Physics, University of Notre Dame, Notre Dame, IN 46556
\end{instit}
\begin{abstract}
Order $\alpha^2$ corrections to the decay rate of
orthopositronium are calculated in the framework of nonrelativistic
QED. The resulting contribution is found to be in significant
disagreement
with one set of experimental measurements, though another experiment is
in agreement with theory.
\end{abstract}

The discrepancy between theory and experiment for the decay rate of
orthopositronium has long been one of the outstanding problems in precision
QED. The
theory is given by
\begin{equation}
\Gamma_{\rm o-Ps} = \Gamma_0 [ 1 + A { \alpha \over \pi} +
{\alpha^2 \over 3} {\rm ln} \alpha + B ({\alpha \over \pi})^2 - {3 \alpha^3
\over 2 \pi}
{\rm ln}^2 \alpha ]
\end{equation}
where the lowest order decay rate is given by
\begin{equation}
\Gamma_0 = {2 \over 9} (\pi^2 - 9) {{m \alpha^6} \over \pi}.
\end{equation}

The value of $\Gamma_0$ is 7.211 169 $\mu s^{-1}$. This is about 2.3
percent above
the experimental determinations at Ann Arbor of
\begin{equation}
\Gamma_{\rm o-Ps}({\rm Gas}) = 7.0514(14) \mu s^{-1} \cite{AnnArborGas}
\end{equation}
\begin{equation}
\Gamma_{\rm o-Ps}({\rm Vacuum}) = 7.0482(16) \mu s^{-1} \cite{AnnArborVac}
\end{equation}
and 2.4 percent above the somewhat less precise Tokyo measurement
\begin{equation}
\Gamma_{\rm o-Ps}({\rm SiO_2}) = 7.0398(29) \mu s^{-1} \cite{Japan}.
\end{equation}
The great bulk of this difference is accounted for by the one-loop
correction \cite{original},
which has been evaluated with high accuracy in \cite{Gregacc} to be
\begin{equation}
A = -10.286~606(10).
\end{equation}
Including this -2.39 percent
effect along with the logarithmic terms of order $\alpha^2 \Gamma_0$
\cite{CL1} and
$\alpha^3 \Gamma_0$ \cite{Karsh}, which contribute -0.01 percent and
-0.0004 percent respectively,
gives a decay rate of 7.038 202 $\mu s^{-1}$. The remaining
difference with the Ann Arbor experiments, with the 2.3 percent difference
reduced to -0.1 percent,
requires a rather large positive value for $B$
($339(36)$ for the gas experiment and $257(41)$ for the vacuum experiment),
which is the
discrepancy mentioned above. The Tokyo experiment, on the other hand, which
disagrees
with the first two by several standard deviations, is consistent with a
small value. While
the experimental situation clearly requires more work, it is also obviously
important to
directly evaluate the constant $B$.

While the need for calculating $B$ has been clear for
two decades, there are two difficulties that have prevented its evaluation
until now. The first is simply the large number, 81, of diagrams that
contribute,
many of which have two-loop ultraviolet infinities. More importantly, a
number of these diagrams have a serious kind of infrared divergence
associated with
the fact that positronium is a bound state.

These problems have recently been overcome for the case of parapositronium
decay \cite{Czar}. The main theoretical tool used in this work is
Nonrelativistic Quantum Electrodynamics (NRQED) \cite{NRQED}.
This approach allows the high-energy part of the problem to be treated
as an on-shell scattering process. In this case the complications of
the bound state are not present, and Feynman gauge can be used. The low-energy,
bound state part can be treated in Coulomb gauge with a small set of operators that
describe relativistic and QED corrections to a Schr\"{o}dinger problem.
A matching procedure carried out with free particle scattering amplitudes
then allows the two parts to be combined into a complete calculation.

The present calculation, while similar in spirit to that of the parapositronium
calculation, regulates long wavelength singularities by giving the photon
a small mass $m_e \lambda$: reference \cite{Czar} instead uses dimensional
regularization. We have chosen to use a slight variation of a recent NRQED 
calculation of one-photon annihilation contributions to ground state positronium 
hyperfine splitting \cite{Hoang} that uses a photon mass, as it is easily generalized 
to the decay rate calculation.

In NRQED the annihilation of orthopositronium can be accounted for by modifying
the amplitude for one-photon annihilation, $2\pi \alpha / m^2$, to
\begin{equation}
V_4 (\vec k, \vec l) = {2 \pi \alpha \over m^2} ( 1 - {4 i \alpha^2 (\pi^2
-9) \over 9 \pi}).
\end{equation}
The independence of $V_4$ on the momentum is a reflection of the fact that
annihilation occurs, on an atomic scale, nearly at a point in coordinate space.
At the level of precision required here we will also need to consider a
modification that accounts for the interaction not being exactly pointlike
\cite{Labelle},
\begin{equation}
V^{\rm der}_4(\vec k, \vec l) = {i \alpha^3 (\pi^2 -9) X \over 27 m^4}
(\vec k^2 + \vec l^2),
\end{equation}
where $X = (19 \pi^2 -132)/(\pi^2 -9)$.

In first-order perturbation theory, taking the expectation value of $V_4$
(which corresponds to multiplying by the square of the wave function
at the origin $m^3 \alpha^3/ 8\pi$) and
using $\Gamma = -2 {\rm Im} (E)$ reproduces Eqn. 2. In addition to this
amplitude
other operators accounting for relativistic effects are present
\cite{Hoang}, and lead to
the following ultraviolet divergent expression in second-order
Rayleigh-Schr\"{o}dinger
perturbation theory, which has the first order effect of $V^{\rm der}_4$
included,
\begin{equation}
\Gamma_{\rm NRQED} = \Gamma_0(1 + {\alpha \over \pi} e_1 +({\alpha \over
\pi})^2 e_2) +
\alpha^2 \Gamma_0 [ -{8 \Lambda \over 3 \pi \alpha } - {1 \over 3}
{\rm ln} {\Lambda \over \alpha} - {11\over 24} - { X \over 12 \pi} ( {2
\Lambda \over \alpha}
- {3 \pi \over 4})].
\end{equation}
Here we have renormalized the imaginary part of $V_4$ with a power series in
$\alpha$ and introduced an ultraviolet
cutoff $m \Lambda$ on the momentum space integrations.

The constants $e_1$ and $e_2$ are determined by requiring that the amplitude
for free particle scattering at threshold in NRQED be equal to that determined
in a complete QED calculation. The one-loop QED calculation at threshold has
an amplitude corresponding to the decay rate
\begin{equation}
\Gamma_1 = { \alpha \over \pi} \Gamma_0 [ {2 \pi \over \lambda} + A(\lambda)]
\end{equation}
where $A(\lambda) = -10.28660 + 15.39 \lambda$. Even though the limit
$\lambda \rightarrow 0$ is taken at the end of the calculation, we keep terms
of order $\lambda$ in the one-loop calculation because some terms enter the
two-loop calculations
with a factor $1/\lambda$. The one-loop matching calculation then allows us
to determine
\begin{equation}
e_1 = { 8 \Lambda \over 3 } + { X \Lambda \over 6} + A(\lambda) -
{ 7 \pi \lambda \over 12} -{ X \pi \lambda \over 12}.
\end{equation}

If we further define the two-loop QED decay rate as
\begin{equation}
\Gamma_2 = ({\alpha \over \pi})^2 \Gamma_0 [ {(2 {\rm ln} 2 + 1) \pi^2
\over \lambda^2} +
{ 2 \pi A(\lambda) \over  \lambda} +{\pi^2 \over 3} {\rm ln} \lambda + B_2 ]
\end{equation}
the two-loop matching calculation gives
\begin{equation}
e_2 = { \pi^2 \over 3} {\rm ln \Lambda} - { \pi^2 X \over 24} + { 11 \pi^2
\over 6} -
{2 \pi^2 {\rm ln} 2 \over 3} + B_2.
\end{equation}
The reason for defining $\Gamma_2$ in terms of $A(\lambda)$ rather than
the physical limit $A(0)$ is a practical one, associated with subtraction
schemes
required to deal with the most infrared divergent two-loop diagrams. It
also has
the advantage of leading to an exact cancellation with the factor
$A(\lambda)$ in $e_1$
in the matching calculation.

With this determination of $e_1$ and $e_2$ the NRQED decay rate becomes
ultraviolet and
infrared finite, and is given by
\begin{equation}
\Gamma_{\rm NRQED} = \Gamma_0 ( 1 + { \alpha \over \pi} A(0) +
 ({\alpha\over \pi})^2 [ {\pi^2 \over 3} {\rm ln} \alpha +
\pi^2 ({11 \over 8} - { 2 {\rm ln} 2 \over 3}) +\pi^2 { X \over 48} + B_2]),
\end{equation}
and numerically the constant we wish to determine is given
by $B = B_2 + \pi^2 (0.9129 + 1.3302) = B_2 + 22.14$. We
note that the constant 0.9129 differs from Ref. \cite{Labelle}, where it is
given as 1.16, the numerical value of 13/8-2/3 ln(2).
An additional contribution of -1/4 to this number has recently been 
found \cite{privateLabelle}, removing the discrepancy. In 
addition Hill and Lepage \cite{Hill-Lepage} have recently
recalculated a number of QED effects in a new nonperturbative implementation of NRQED, 
and obtain 0.9125(5), so all NRQED calculations are now in agreement. As
an additional check of the method, we verified that our implementation of NRQED,
when applied to the one-photon
annihilation contribution to ground state hyperfine splitting, reproduces the
known answer \cite{Hoang}. 

To finish the calculation the free two-loop QED calculation must be carried
out and $B_2$ extracted.  While the two-loop calculation is sufficiently involved 
that we must defer its detailed description to a longer work \cite{future}, it is 
useful to refer to the diagrams that enter the one-loop calculation. The six diagrams 
of Fig. 1 we refer to as the outer vertex (OV), inner vertex (IV), double vertex (DV), 
self-energy (SE), ladder (L), and annihilation (A) contributions. After the 
ultraviolet divergences are removed by renormalization,
the individual values of the diagrams are presented in Table I.

The 81 QED diagrams that contribute to the decay rate at two-loop order
break into 11 classes that we label a-k.  Class a consists  of irreducible 
two-loop vertex corrections which generalize the OV diagrams. While free of 
binding  singularities, their evaluation is complicated by the need to carry 
out two-loop renormalization.  Similar comments apply to Class b, the two-loop 
generalization of the IV diagram, and Class c, the generalization of
the SE diagram. We regulate ultraviolet divergences by using  
dimensional regularization, working in $n=4-2 \epsilon$ dimensions, and
the finite photon mass regulates infrared divergences.
Renormalization constants in this scheme have not to our knowledge been presented in the
literature: details of their calculation will be given elsewhere \cite{GregFell}.

Class d consist of diagrams with reducible two-loop
corrections, in which two separated ultraviolet divergent one-loop corrections 
are present. The next set of diagrams, which have no ultraviolet or
infrared singularities, are those of class e, which generalize the DV diagrams. 
The most difficult to evaluate diagrams were in the
f class, which generalize the L diagram. Most of these contributions
diverge as $1/\lambda$, and the most singular as $1/\lambda^2$. 
Canceling ln$\lambda/\lambda$ divergences characteristic of Feynman
gauge were present that were quite difficult to handle numerically.

Class g consists of 9 diagrams in which the DV diagram has ultraviolet
divergent one-loop radiative corrections in all possible vertices and
propagators.  Class h consists of radiative corrections to the A diagram, 
and have previously been calculated in Ref. \cite{Lymb}. Because that
calculation used a Bethe-Salpeter formalism a ln$\alpha$ was present that
has to be replaced with a ln$\lambda$ in our present formalism: the additive
constant however is unchanged.  Class i consists of diagrams where a vacuum
polarization loop has been inserted in all possible places in the one-loop 
calculation.  These have also been previously treated \cite{BurVP} and \cite{VP},
as have the last two-loop effect we include, the square of the one-loop
amplitude, which we call class j \cite{Bur}, \cite{Gregacc}. As with class h,
our present formalism leads to terms that depend on $\lambda$, but the
additive constant is again unchanged.

Finally, class k, which involves two of the three photons emitted in the decay
undergoing light-by-light scattering have not been calculated:
because of the small numerical contributions of these diagrams in
parapositronium \cite{Czar} we consider it
highly unlikely that the omission of these diagrams will affect our
conclusions.

The results of the calculation are summarized in Table II.

\vskip 2.0cm

\mediumtext
\begin{table}
\caption{Renormalized one-loop contributions to the orthopositronium decay
rate.}
\label{table_1R}
\begin{tabular}{cccdd}
contribution
& ${\pi \over \lambda} \bigl ( {\alpha \over \pi} \bigr ) \Gamma_0$
& $\ln\lambda \bigl ( {\alpha \over \pi} \bigr ) \Gamma_0$ & ${\alpha
\over \pi} \Gamma_0$ & $\lambda {\alpha \over \pi} \Gamma_0$ \\
\tableline
$\Gamma_{\rm OV}$ & 0 & -4 & -1.028861425 &  1.8756(1) \\
$\Gamma_{\rm IV}$ & 0 & -2 & -1.839322925 &  4.7124    \\
$\Gamma_{\rm DV}$ & 0 & 0 & -3.567629(21) &  7.5499(2) \\
$\Gamma_{\rm SE}$ & 0 & 4 & 4.784983909   &-11.0445(1) \\
$\Gamma_{\rm L}$  & 2 & 2 & -7.821768(32) & 12.296(4)  \\
$\Gamma_{\rm A}$  & 0 & 0 & -0.8140573(3) &  0.0       \\
\tableline
total & 2 & 0 & -10.286606(10) &  15.389(4)            \\
\end{tabular}
\end{table}

\vskip 2.0cm

\mediumtext
\begin{table}
\caption{Contributions to the orthopositronium decay rate by class.}
\label{table_2}
\begin{tabular}{ccccccd}
diagram &
${\alpha^2 \over \lambda^2} \Gamma_0$                &
${\alpha^2 \over \pi \lambda} \Gamma_0$              &
$\ln\lambda {\alpha \over \pi}$                      &
$({\alpha \over \pi})^2 {\rm ln}^2 \lambda \Gamma_0$ &
$({\alpha \over \pi})^2 {\rm ln} \lambda \Gamma_0$   &
$({\alpha \over \pi})^2 \Gamma_0$                   \\
\tableline
a  & 0       &  0 & $-\Gamma_{\rm OV}$         &-2 & 0           &  -5.618
\\
b  & 0       &  0 & $-\Gamma_{\rm IV}$         &-1 & 0           &  -0.705
\\
c  & 0       &  0 & $-\Gamma_{\rm SE}$         & 2 & 0           &   0.058
\\
d  & 0       &  0 & 0                          & 0 & 0           &   2.421
\\
e  & 0       &  0 & 0                          & 0 & 0           &
9.259(9)  \\
f  & 2 ln 2  &  A & $\Gamma_{\rm SE+OV+IV+DV}$ & 1 & $-2\pi^2/3$ &
-20.50(26)  \\
g  & 0       &  0 & $-\Gamma_{\rm DV}$         & 0 & 0           &  -1.372
\\
h  & 0       &  0 & 0                          & 0 & $\pi^2$     &   9.007
\\
i  & 0       &  0 & 0                          & 0 & 0           &   0.965
\\
j  & 1       &  A & 0                          & 0 & 0           &  28.860(2)
\\
\tableline
total&2ln 2+1&  2A& 0                          & 0 & $\pi^2/3$   & 22.38(26) \\
\end{tabular}
\end{table}

We see from Table II that $B_2 = 22.38(26)$, so the complete result for $B$ is
44.52(26). While this is indeed a relatively large contribution, it leaves the theoretical
prediction,
\begin{equation}
\Gamma = 7.039~934(10) \mu s^{-1}
\end{equation}
well below the Ann Arbor results, by 8 and 5 standard deviations for the gas and
vacuum measurements respectively, though consistent with the Tokyo measurement.

No conclusions can be drawn until the experimental situation is clarified.
If the Ann Arbor results are confirmed, while of course it is tempting to consider explaining
this effect through exotic interactions \cite{exotic}, it is worth noting
that there is also at present a significant discrepancy between theory and experiment
in the ground state hyperfine splitting of positronium. While there were disagreements
between various calculations for some time, recently complete agreement \cite{Pach},\cite{Czar2},
\cite{Erratum}, \cite{Hill-Lepage} on a value of 203 392.05 MHz has been found. This 
value is 4 experimental standard deviations above the Yale measurement \cite{Yale}
and 2.8 above the Brandeis measurement \cite{Brandeis}. If the present disagreement
of theory and experiment in positronium persists, any explanation in terms of new physics
would be most compelling if both discrepancies were explained by the same mechanism.

The work of JS was partially supported by NSF grant PHY-9870017, and that
of GA by NSF grants PHY-9711991 and PHY-9722074. GA acknowledges the hospitality
of the Aspen Center for Theoretical Physics and the Physics department of UCLA.
We thank P. Labelle, G.P. Lepage  and R. Hill for very useful conversations, and the 
constant encouragement of the Michigan experimental group is also gratefully acknowledged.

Figure captions:

Fig. 1 Diagrams contributing to the one-loop decay rate correction.

\begin{references}
\bibitem{AnnArborGas} C.I. Westbrook, D.W. Gidley, R.S. Conti, and A. Rich,
Phys. Rev. A {\bf 40}, 5489 (1989).
\bibitem{AnnArborVac} J.S. Nico, D.W. Gidley, A. Rich, and P.W. Zitzewitz,
Phys.
Rev. Lett {\bf 65}, 1344 (1990).
\bibitem{Japan} S. Asai, S. Orito, and N. Shinohara, Phys. Lett. B {\bf
357}, 475 (1995).
\bibitem{original} W.E. Caswell, G.P. Lepage, and J. Sapirstein, Phys. Rev.
Lett. {\bf 38},
488 (1977).
\bibitem{Gregacc} G.S. Adkins, Phys. Rev. Lett. {\bf 76}, 4903 (1996).
\bibitem{CL1} W.E. Caswell and G.P. Lepage, Phys. A {\bf 20}, 36 (1979).
\bibitem{Karsh} S.G. Karshenboim, Zh. Eksp. Teor. Fiz. {\bf 103}, 1105
(1993) [JETP {\bf 76},
541 (1993)].
\bibitem{Czar} A. Czarnecki, K. Melnikov, and A. Yelkhovsky, Phys. Rev.
Lett. {\bf 83},
1135 (1999) and preprint hep-ph/9910488.
\bibitem{NRQED} W.E. Caswell and G.P. Lepage, Phys. Lett. {\bf 167B}, 437
(1986).
\bibitem{Hoang} A.H. Hoang, P. Labelle, and S.M. Zebarjad, hep-ph/9909495.
\bibitem{Labelle} P. Labelle, G.P. Lepage, and U. Magnea, Phys. Rev. Lett.
{\bf 72}, 2006 (1994).
\bibitem{privateLabelle} P. Labelle, private communication.
\bibitem{Hill-Lepage} R. Hill and G.P. Lepage, Cornell preprint (2000) and private 
communication.
\bibitem{future} G.S. Adkins, R.N. Fell, and J. Sapirstein, in preparation.
\bibitem{GregFell} G.S. Adkins, R.N. Fell, and J. Sapirstein, in preparation.
\bibitem{Lymb} G.S. Adkins and M.  Lymberopoulos, Phys. Rev. A {\bf 51},
2908 (1995).
\bibitem{BurVP} A.P. Burichenko and D. Yu. Ivanov, Phys. At. Nucl. {\bf
58}, 832 (1995) [Yad. Fiz. {\bf 58}, 898 (1995)].
\bibitem{VP} G.S. Adkins and Y. Shiferaw, Phys. Rev. A {\bf 52}, 2442 (1995).
\bibitem{Bur} A.P. Burichenko, Phys. At. Nucl. {\bf 56}, 640 (1993) 
[Yad. Fiz. {\bf 56}, 123 (1993)].
\bibitem{exotic} A. Czarnecki and S.G. Karshenboim, hep-ph/9911410.
\bibitem{Pach} K. Pachucki, Phys. Rev. A {\bf 56}, 297 (1997).
\bibitem{Czar2} A. Czarnecki, K. Melnikov, and A. Yelkhovsky, Phys. Rev. Lett. {\bf 82},
311 (1999); Phys. Rev. A {\bf 59}, 4316 (1999).
\bibitem{Erratum} G. Adkins and J. Sapirstein, erratum to Phys. Rev. A {\bf 58}, 3552 (1998).
\bibitem{Yale} M.W. Ritter, P.O. Egan, V.W. Hughes, and K.A. Woodle, Phys. Rev.
A {\bf 30}, 1331 (1984).
\bibitem{Brandeis} A.P. Mills, Jr., Phys. Rev. A {\bf 27}, 262 (1983).
\end{references}
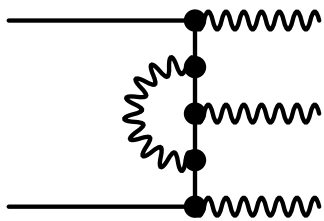
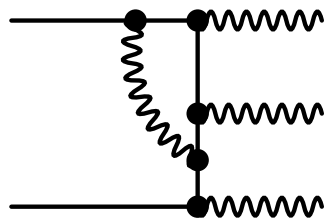
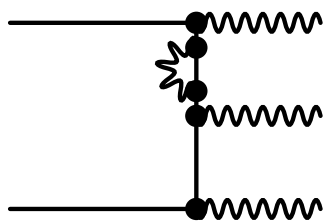
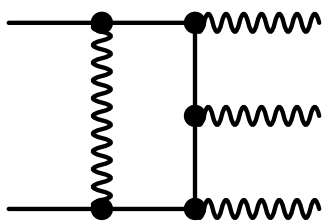
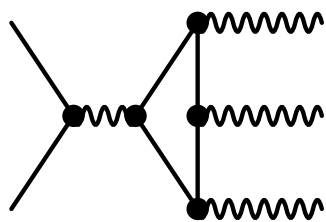
\end{document}